\documentclass[english]{cccconf}
\usepackage[comma,numbers,square,sort&compress]{natbib}
\usepackage{amsmath}
\begin{document}

\title{A Unified Stability Analysis Approach for A Class of Interconnected System}
\author{WANG Yong}



\affiliation{Beijing Institute of Control Engineering,
         Beijing 100190, P.~R.~China\email{wyongzju@163.com}}


\maketitle

\begin{abstract}
From the structural perspective, this paper investigates a new formulation of the concept of input-to-state stability (ISS), and based on this formulation, proposes a new stability analysis approach for a class of interconnected system. The new formulation of ISS is better able to reflect the tendency of the state $x(t)$  tracking the input  $u(t)$ and weakens the conservative of the original form. The stability analysis method which transforms the interconnected system into the equivalent cascade form, does not depend on the Lyapunov function, breaks through the limitation of the small-gain theorem and extends the application of  ISS. As its applications in  three typical kinds of interconnected systems, this method is used to prove the small-gain theorem again and analyzes the stability of a class of interconnected system and the consensus of the multi-agent system (MAS).
\end{abstract}

\keywords{ input-to-state stability, interconnected system, cascade system, small-gain theorem, stability analysis, multi-agent system..}

\footnotetext{This research is supported by the National Natural Science Foundation of China under grant 61333008 and the National Basic Research
Program (973) of China under Grant 2013CB733100.}

\section{Introduction}
\subsection{ Background}
The concept of the input-to-state stability (ISS) was introduced by E.D.Sontag in 1989 in the well-known paper\cite {E.D.Sontag1}, which becomes a popular method to study the input-output property of nonlinear systems later. Generally, a system $\dot x = f\left( {t,x,u} \right)$ is said to be ISS if there exist  class $KL$ function $\beta$ and  class $K$ function $\gamma$ , such that for any initial value in the closed set $D$ and the bounded input $u(t)$ , the following is satisfied:
\[\left\| {x(t)} \right\| \le \beta (\left\| {{x_0}} \right\|,t - {t_0}) + \gamma (\mathop {\sup }\limits_{\tau  \in [{t_0},t)} \left\| {u(\tau )} \right\|)\]
ISS describes the evolution of  the state of a stable system when it is driven by the external input.  Based on the concept,  many researchers proposed some other concepts, such as IOS(input to output stable),OLIOS(output-Lagrange input to output stable),SIIOS(state-independent IOS), ROS(robustly output stable)\cite{E.D.Sontag2},iISS (integral input-to-state stable)\cite{Z.Jiang1} , and ISDS (input-to-state dynamically stable) \cite{S. N. Dashkovskiy} which describes the dynamic process of a stable system. The introduction of  ISS gives us a new way to study the stability of a system. For more complicated systems, which contain many subsystems that interconnect with each other, if every subsystem is ISS, we can take a structural  perspective, ignore their internal details and take full advantage of  the input-output property and interconnected relationship between subsystems to study the stability problem. In this way, we only need to focus on the relationship between all subsystems and need not care about all details,which is greatly different from the Lyapunov function based method. Therefore, the concept of ISS greatly simplifies the stability analysis of the complex interconnected system.

A good example of the above idea is the well-known small-gain theorem. In short, if  two subsystems are ISS and interconnected with each other, if  the composition of the gain function (the quantitative expression of the input-output property of each subsystem)along the closed cycle is less than the identity function, the entire system is stable. Later on, with ISS as tool, many researchers extended the small-gain theorem from linear systems to nonlinear systems and proposed its various forms and the associated proofs \cite{S. Dashkovskiy2}-\cite{A.R.Teel}. Furthermore, Jiang and etc. extend the small-gain theorem to the case of the interconnected system with more than two subsystems, and give the sufficient condition that ensures stability in \cite {Z. Jiang2}, i.e., if the composition of the gain function along every closed cycle is less than the identity function, the entire system is stable. Because the structural perspective  is brief and intuitionistic,  it becomes an important method to design the controller  \cite{Z.Jiang3}\cite{R. Marino1}and analyze the stability .

However, when facing some new problems, the above concept and approach meet across some difficulties .

 {\emph{\emph{ 1)~~The form of ISS is too conservative to describe the
corresponding change of the state $x(t)$ of system tracking the input $u(t)$.}}}

 $\mathop {\sup }\limits_{\tau  \in [{t_0},t)} \left\| {u(\tau )} \right\|$ just denotes the maximum of $u(t)$  in a certain time interval, but in fact, $x(t)$  keeps tracking the change of the input $u(t)$ all the time so as to be kept in a neighboring area of it. Therefore $\mathop {\sup }\limits_{\tau  \in [{t_0},t)} \left\| {u(\tau )} \right\|$  is too conservative to describe the fact, especially, when $u(t)$ converges towards a constant.

 2)~~Some systems, only one of whose subsystems is ISS, are still stable in fact.

   The small-gain theorem requires all subsystems should be ISS, but in fact, some systems like the one given by equation (1) below, are also stable even though the first subsystem is not ISS.

\emph{\textbf{Example 1.}}
\begin{equation}  \label{1}
\left\{ \begin{array}{l}
\dot x = z\\
\dot z =  - z - x
\end{array} \right.
\end{equation}
where $x,z \in R$. The z-subsystem,i.e. the second equation, is ISS, but the x-subsystem,i.e. the first equation, is not.

 3) Even though  every agent is ISS, the consensus problem of  multi-agent systems is not explained via the small-gain theorem.Since the composition of the gain function along the closed cycle equals the identity function,which conflicts with the small-gain theorem.

 Therefore, from the structural perspective, the tools in the current literature can not solve these new problems, which urgently requires to develop a new formulation of ISS and a new approach.

\subsection{ Problem Statement}

In this paper, we investigate a new formulation of ISS, and based on it, propose a stability analysis approach for a class of interconnected system. The contribution of this paper includes three parts. Firstly, a new formulation of ISS is proposed, which is better able to reflect the corresponding change of $x(t)$  tracking $u(t)$  than the original form, and using this new formulation, we uncover the essential relationship between the interconnected system and the cascade system. Secondly, based on the research of ISS, a unified stability analysis approach is proposed, which does not depend on the construction of a Lyapunov function. Thirdly, as the application of this approach, we analyze the stability of three typical kinds of interconnected systems, i.e., repeating to prove  the small-gain theorem in this new framework and analyzing the stability of a class of interconnected system like equation (1) and the consensus of  a multi-agent system.

\subsection {Organization of the Paper}

The rest of paper is organized as follows. Section 2 gives some notations and briefly recalls some basic background knowledge. Section 3 presents the framework of this approach. Section 4 presents a new formulation of ISS. Section 5 gives its applications for three typical interconnected systems. Section 6 is the conclusion and talks about other applications.

\section{Notation and Preliminaries}

Classes of $K,{K_\infty },KL$  and positive definite function follow the definition in [10], which are extensively used in the field.$D \in R^n$ denotes a domain containing the origin.
 Now , we recall the traditional concept of ISS again. Consider the general nonlinear system
\begin{equation}  \label{2}
\dot x = f(x,u)
\end{equation}
where $f:{R^n} \times {R^m} \to {R^n}$ is the continuous function and local Lipschitz function w.r.t.$x(t)$  and $u(t)$ .


The Lyapunov-like theorem that follows gives a sufficient and necessary condition for ISS.

\textbf{\emph{Proposition 1.}}[10]  Let $V:[0,\infty ] \times {R^n} \to R$  be a continuously differentiable function such that
\begin{equation}  \label{4}
{\alpha _1}(\left\| x \right\|) \le V \le {\alpha _2}(\left\| x \right\|)
\end{equation}
\begin{equation}  \label{5}
\dot V \le  - {W_3}(x),\forall \left\| x \right\| \ge \rho (\left\| u \right\|) > 0
\end{equation}
where ${\alpha _1},{\alpha _2} \in {K_\infty }$ ,  $\rho  \in K$,and ${W_3}$  is continuous positive definite function on  ${R^n}$.Then ,the system (\ref{2}) is input-to-state stable with $\gamma  = \alpha _1^{ - 1} \circ {\alpha _2} \circ \rho $ .

\section{Problem Statement and Analysis Framework}

In this paper, we investigate the stability problem of a class of  interconnected system described by the following

Subsystem $x$:
\begin{equation}  \label{6}
\dot x = {f_1}(x,z),
\end{equation}

Subsystem $z$:
\begin{equation}  \label{7}
\dot z = {f_2}(z,x),
\end{equation}
where  $x \in {R^{{n_1}}}$,$z \in {R^{{n_2}}}$,${f_1}$ and ${f_2}$ are   continuous functions. Assume that at least one subsystem is ISS, and without loss of generality, suppose x-subsystem is ISS.

Naturally, Lyapunov function is the most general choice for the stability analysis of such  system. However, it is not  easy to find an appropriate candidate function, especially when the system becomes more and more complex and a lot of subsystems are strongly coupled with each other. If  every subsystem is ISS, we can resort to the small-gain theorem for analysis, which uses the gain of each subsystem  to check the stability of the interconnected system. Essentially, such a way represents a structural perspective and is more suitable for interconnected systems than the Lyapunov function based approach.
Inspired by this idea, this paper will propose a new structural stability analysis approach  for (\ref{6}) and (\ref{7}) in the following.

\emph{\textbf{Stability Analysis Procedure 1.}}

 \textbf{Step 1} ~~Transform the interconnected form into a cascade  form via the ISS property of x-subsystem .

  The solution of x-subsystem can be written as a function of  the initial value $x({t_0})$ , $t$ and input  $u(t)$ , i.e.
  $x(t){\rm{ = }}\phi ({x_0},t,z)$.
Substituting this equation into the z-subsystem yields
$\dot z = {f_2}(z,\phi ({x_0},t,z))$.
Thus, the interconnected system becomes a cascade system  of the following form
\begin{equation}  \label{9}
\dot z = {f_2}(z,\phi ({x_0},t,z))
\end{equation}
\begin{equation}  \label{10}
\dot x = {f_1}(x,z)
\end{equation}

\textbf{\emph{Remark 1.}}
Using the ISS property, the  above process has an intuitive explanation.  Since the x-subsystem  is ISS w.r.t. $z(t)$  and suppose the gain function from $z(t)$   to
 $x(t)$ is $\gamma$ , let $\gamma (z(t))$ be the input, then the x-subsystem corresponds to a filter whose gain is 1. Its output is not arbitrary but keeps tracking  $\gamma (z(t))$ , and in fact, it is kept in a neighboring area of  $\gamma (z(t))$ . Therefore,  equation (8) can be written as the following formulation of ISS
 \begin{equation}  \label{11}
x(t) = \beta ({x_0},t) + \gamma (z(t)) + \Delta
\end{equation}
 where $\Delta$ denotes  a static error with $\gamma (z(t))$  whose specific form will be given later. Therefore, using (\ref{11}), we can construct a feedback loop as follows
 \begin{equation}  \label{12}
\dot z = {f_2}(z,\beta ({x_0},t) + \gamma (z) + \Delta )
\end{equation}
and the cascade system can be described in the Fig.1.
\begin{figure}
\includegraphics[scale=0.6 ]{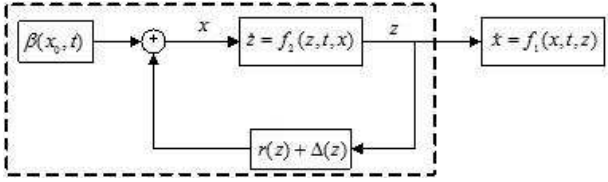}
\caption{The cascade form of the interconnected system}\label{fig:side:a}
\end{figure}

 \textbf{Step 2.}~~Analyze the stability of  the feedback loop of z-subsystem.

Since $\beta ({x_0},t)$ is convergent, the stability of feedback loop of z-subsystem depends on the function ${f_2}$ and $\gamma(z) + \Delta $ . It is necessary to study the  new formulation of ISS.

\textbf{\emph{Remark 2.}}
Compare with the original formulation of  ISS, $\gamma(z) + \Delta $  replaces  $\gamma (\mathop {\sup }\limits_{\tau  \in [{t_0},t)} \left\| {u(\tau )} \right\|)$ . It should note that the former represents the current value of the input $z(t)$,but the later represents its history. Besides, in various proofs of the small-gain theorems(\cite{S. Dashkovskiy2}-\cite{A.R.Teel}), only the form of the gain  $\gamma$ is needed, but
 $\Delta $ is not cared absolutely. But later, we will show what is $\Delta $ and what does it function in the new problem.

 \textbf{Step 3}~~Analyze the stability of  the cascade system.

After transformed into the cascade system,  by the stability theorems in [11]  about cascade systems and the ISS property of  x-subsystem ,  if and only if  the z-subsystem is stable,  the entire system is stable, so is the original interconnected system.

  \textbf{\emph{Remark 3.}}
  In this approach, the concept of ISS bridges the gap of the interconnected system and the cascade system. In this way, the stability of the complex interconnected system is equivalent to the stability of its equivalent cascade system, and further the stability of a feedback loop. Through transforming the stability of the interconnected system into the stability of one of subsystems, this approach greatly simplifies the analysis.

It should be mentioned that this approach just requires one subsystem should be ISS and need not construct an overall Lyapunov function which used to consider all details and too depends on the specific form of the system.

\section{New Formulation of ISS}
In this section, we present a new formulation of ISS concept to weaken the conservative of the original one.
  Consider the general system  (\ref{2}) and suppose it satisfies the following assumption.

\textbf{\emph{Assumption 1.}} The general system  (\ref{2}) satisfies the proposition 1 and the gain function $\gamma  = \alpha _1^{ - 1} \circ {\alpha _2} \circ \rho $ is differentiable.

Then we have the new formulation of ISS in the following.

\textbf{\emph{Theorem 1.}}~~Suppose the system (\ref{2}) satisfies the assumption 1 in $D$, if there exist a class $KL$ function   $\beta$,  a class $K$ function $\gamma$, and a  constant $L > 0$,such that for any initial state $x({t_0})$  and any bounded input $u(t)$ , the solution of $x(t)$ exists for all $t \ge {t_0}$  and satisfies
\[\left\| x \right\| \le \beta ({x_0},{u_0},t) + \gamma (\left\| u \right\|) - L\int_0^t {{e^{ - \int_s^t {k(\tau )d\tau } }}} {\alpha _4}(u)\dot uds\]
where ${\alpha _4}(u) = \frac{{d{\alpha _2}(\rho (\left\| u \right\|))}}{{du}}$  and $k(t)$ is continuous positive definite on $R$ .

\emph{proof.}
By the proposition 1, there exists the Lyapunov function $V(x)$ such that
\begin{equation}  \label{13}
{\alpha _1}(\left\| x \right\|) \le V \le {\alpha _2}(\left\| x \right\|)
\end{equation}
\begin{equation}  \label{14}
\dot V \le  - {W}(V),\forall \left\| x \right\| \ge \rho (\left\| u \right\|) > 0.
\end{equation}
where $W$ is positive definite  on $R$, and $W(V)$
can be obtained by  (\ref{13}).
According to (\ref{13}), the condition of (\ref{14}) $\left\| x \right\| \ge \rho (\left\| u \right\|)$ can be strengthened as $V \ge {\alpha _2}(\rho (\left\| u \right\|))$ , and define the error $e = V - {\alpha _2}(\rho (\left\| u \right\|))$ ,then  we obtain the error system of equation (\ref{14})
\begin{equation}  \label{15}
\dot e \le  - {W}(e + {\alpha _2}(\rho (\left\| u \right\|))) - {\alpha _4}(u)\dot u,\forall e \geq 0
\end{equation}
where  ${\alpha _4}(u) = \frac{{d{\alpha _2}(\rho (\left\| u \right\|))}}{{du}}$.
Since ${W} $ is positive definite, we have
$k(t) = \frac{{{W}(e + {\alpha _2}(\rho (\left\| u \right\|)))}}{e} > 0,\forall e > 0$.
Especially,$e = 0$ means $\dot u = 0$ and $\left\| x \right\| \le \gamma (\left\| u \right\|)$.
Therefore, (\ref{15}) can be written as
\begin{equation}  \label{16}
\dot e \le  - k(t)e - {\alpha _4}(u)\dot u.
\end{equation}
Solving it and by the comparison theorem in [10]    yields
\begin{equation}  \label{17}
e(t) \le {e^{ - \int_0^t {k(s)ds} }}{e_0} - \int_0^t {{e^{ - \int_s^t {k(\tau )d\tau } }}} {\alpha _4}(u)\dot uds
\end{equation}
Due to $\int_0^t {k(s)ds}  > 0$ , we  define the  class $KL$ function
${\beta _1}({x_0},{u_0},t) = {e^{ - \int_0^t {k(s)ds} }}{e_0}$.
 In view of $e = V - {\alpha _2}(\rho (\left\| u \right\|))$ ,equation (\ref{17}) can be written as
\begin{equation}  \label{18}
V(t) \le {\beta _1}({x_0},{u_0},t) + {\alpha _2}(\rho (\left\| u \right\|)) - \int_0^t {{e^{ - \int_s^t {k(\tau )d\tau } }}} {\alpha _4}(u)\dot uds.
\end{equation}
That is
\[\left\| x \right\| \le \alpha _1^{ - 1}({\beta _1}({x_0},{u_0},t) + {\alpha _2}(\rho (\left\| u \right\|)) - \int_0^t {{e^{ - \int_s^t {k(\tau )d\tau } }}} {\alpha _4}(u)\dot uds).\]
By the Lagrange median theorem, we have
\begin{align} \label{20}
 &\alpha _1^{ - 1}({\beta _1}({x_0},{u_0},t) + {\alpha _2}(\rho (\left\| u \right\|)) - \int_0^t {{e^{ - \int_s^t {k(\tau )d\tau } }}} {\alpha _4}(u)\dot uds) \nonumber\\
 & - \alpha _1^{ - 1}({\alpha _2}(\rho (\left\| u \right\|)) \nonumber\\
&= \frac{{d\alpha _1^{ - 1}(x)}}{{dx}}\left| {_{x = \xi }} \right.({\beta _1}({x_0},{u_0},t) - \int_0^t {{e^{ - \int_s^t {k(\tau )d\tau } }}} {\alpha _4}(u)\dot uds)
 \end{align}
where $x$ denotes ${\alpha _2}(\rho (\left\| u \right\|))$.
Due to $\alpha _1^{ - 1}(x) \in K$ and $\frac{{d\alpha _1^{ - 1}(x)}}{{dx}} > 0$ , for $x \in D$ , there exists a constant $L>0$, such that
\begin{align}  \label{21}
\left\| x \right\| &\le \alpha _1^{ - 1}({\alpha _2}(\rho (\left\| u \right\|)) + L{\beta _1}({x_0},{u_0},t)\nonumber\\
 &- L\int_0^t {{e^{ - \int_s^t {k(\tau )d\tau } }}} {\alpha _4}(u)\dot uds .
\end{align}
At last, equation (\ref{21}) can be written as
\begin{align}  \label{22}
\left\| x \right\| \le \beta ({x_0},{u_0},t) + \gamma (\left\| u \right\|) - L\int_0^t {{e^{ - \int_s^t {k(\tau )d\tau } }}} {\alpha _4}(u)\dot uds
\end{align}
where $\beta  = L{\beta _1}$,$\gamma  = \alpha _1^{ - 1} \circ {\alpha _2} \circ \rho $.

\textbf{\emph{Remark 4.}}
Compare with the old form, $\gamma (\left\| u \right\|) - L\int_0^t {{e^{ - \int_s^t {k(\tau )d\tau } }}} {\alpha _4}(u)\dot uds$ is more accurate than $\gamma (\mathop {\sup }\limits_{\tau  \in [{t_0},t)} \left\| {u(\tau )} \right\|)$ . Recall equation (\ref{11}),$\Delta  =  - L\int_0^t {{e^{ - \int_s^t {k(\tau )d\tau } }}} {\alpha _4}(u)\dot uds$, and
its convergence depends on the existence of  $\mathop {\lim }\limits_{t \to \infty } \int_0^t {{e^{ - \int_s^t {k(\tau )d\tau } }}} {\alpha _4}(u)\dot uds$.

\textbf{\emph{Remark 5.}}
The new formulation reflects the tendency of $x(t)$  tracking $u(t)$. If ignore the influence of  $x({t_0})$, the neighboring area  is determined by $\Delta $ , that is when $\dot u \to 0$ ,$\Delta  \to 0$ ,then $\left\| {x(t)} \right\| \le \gamma (\left\| u \right\|)$ or  $\left\| {x(t)} \right\| \to \gamma (\left\| u \right\|)$. When $\dot u$ is bounded and $\Delta $ is exists,  $x(t)$  keeps in a specific neighboring area of   $u(t)$. When $\dot u \to \infty $ or $\Delta $ does not exists, the neighboring area is boundless. In other words, if we treat $u(t)$ as the leader and $x(t)$ as the follower, ${\alpha _4}(u)\dot u$ denotes the change rate of the leader, while ${e^{ - \int_s^t {k(\tau )d\tau } }}$ stands for the tracking rate of the follower.

Based on the theorem 1, we have another formulation of ISS in the following.

\textbf{\emph{Corollary  1.}}~~Suppose the condition of theorem 1 is satisfied, the solution of system (\ref{2}) can be written as  .
\begin{equation}  \label{23}
\left\| x \right\| \le \beta ({x_0},{u_0},t) + \alpha _1^{ - 1}(({\rm{1}} - {e^{ - \int_\xi ^t {k(s)ds} }}){\alpha _2}(\rho (\left\| u \right\|)))
\end{equation}
where $\xi  \in [0,t]$ and $k(t)$ is continuous positive definite  on $R$.

\emph{proof.}
By the integral median theorem,  $- \int_0^t {{e^{\int_s^t {k(\tau )d\tau } }}} {\alpha _4}(u)\dot uds$ in equation (\ref{17}) can be written as
\begin{align}\label{24}
  &- \int_0^t {{e^{\int_s^t {k(\tau )d\tau } }}} {\alpha _4}(u)\dot uds  \nonumber\\
 &=  - {e^{ - \int_\xi ^t {k(s)ds} }}\int_0^t {{\alpha _4}(u)\dot uds}  \nonumber\\
 &=  - {e^{ - \int_\xi ^t {k(s)ds} }}{\alpha _2}(\rho (\left\| u \right\|)){\rm{ + }}{e^{ - \int_\xi ^t {k(s)ds} }}{\alpha _2}(\rho (\left\| {u({t_0})} \right\|)) 
\end{align}
where $\xi  \in [{\rm{0,}}t]$.
So  equation(\ref{18}) can be written as
\begin{align}\label{25}
V & \le \beta_1 ({x_0},{u_0},t) + {\alpha _2}(\rho (\left\| u \right\|)) - {e^{ - \int_\xi ^t {k(s)ds} }}{\alpha _2}(\rho (\left\| u \right\|))\nonumber\\
 & + {e^{ - \int_\xi ^t {k(s)ds} }}{\alpha _2}(\rho (\left\| {u({t_0})} \right\|))
\end{align}
Since ${e^{ - \int_\xi ^t {k(s)ds} }} \in (0,1)$, we have
\begin{equation}  \label{26}
V \le {\beta _2}({x_0},{u_0},t) + ({\rm{1}} - {e^{ - \int_\xi ^t {k(s)ds} }}){\alpha _2}(\rho (\left\| u \right\|))
\end{equation}
where $\beta_2 ({x_0},{u_0},t){\rm{ = }}{\beta _1}({x_0},{u_0},t) + {e^{ - \int_\xi ^t {k(s)ds} }}{\alpha _2}(\rho (\left\| {u({t_0})} \right\|))$.

Following the way of theorem 1, for (\label{26}) we obtain
\[\left\| x \right\| \le \beta ({x_0},{u_0},t) + \alpha _1^{ - 1}{\rm{((1}} - {e^{ - \int_\xi ^t {k(s)ds} }}){\alpha _2}(\rho (\left\| u \right\|)))\]~~~

\textbf{\emph{Remark 6.}}
Compare two formulations (19) and (20), the formulation  (19)  containing $\dot u$ is suitable for the analysis of the consensus of  multi-agent systems, while (20) only containing  $ u$ is convenient for the general stable system . 

The following simple example can illustrate the theorem 1.

Now ,let us consider a simple case. Let $u \in R$  in the system (\ref{2}), we obtain the system that follows
\begin{equation}  \label{27}
\dot x = f(x,u)
\end{equation}
It satisfies the assumption.

\emph{\textbf{Assumption 2.}}
System (\ref{27}) is ISS, and there exists differentiable function $\gamma :R^n \to {R^n}$ whose element ${\gamma _i}$ belongs to the class $K$ function,  such that
\begin{equation}  \label{28}
0 = f(\gamma (u),u)
\end{equation}

\textbf{\emph{Remark 7.}}
The assumption represents a class of system whose equilibrium point is a class $K$ function of input $u(t)$,  that means its static gain can be obtained by solving the algebraic equation (\ref{28}). There are many examples, e.g. all linear systems and  the following nonlinear systems $\dot x =  - {x^3} + u$ and $\dot x =  - \tan x + \tan u$ .

Before moving on, we introduce a lemma first.

\textbf{\emph{Lemma 1.}} If the function matrix $A(x) = \{ {a_{ij}}(x)\} $,${a_{ij}}:{R^n} \to R$ ,$x\in D$,$\forall i ,j = 1,...,n$,
 is negative (or positive) definite,  there exist scalar functions ${\lambda _1}(x) < {\lambda _2}(x) < 0$ (or ${\lambda _2}(x) > {\lambda _1}(x) > 0$ )such that
\[{\lambda _1}(x)I < A(x) < {\lambda _2}(x)I.\]
\emph{Proof.} Since for any square matrix  $A = \{ {a_{ij}}\} $,$\forall i ,j = 1,...,n$ ,there exists a non-singular matrix $P$ and a Jordan form $J = diag({J_i})$ , $\forall i = 1,...,m,m \le n$ ,such that
$A = PJ{P^{ - 1}}$.
Then for $A(x)$, by the continuity,  we have
\begin{equation}  \label{29}
A(x) = P(x)J(x){P^{ - 1}}(x).
\end{equation}
When $A(x)$ is negative definite in $D$, its every engenvalue ${\lambda _{i,j}}(x)$ of ${J_i}$ is negative,where $j=1,...,rank(J_i)$.

Then by the continuity, there exists $\lambda_{min} (x) < \lambda_{max} (x) < 0$ such that $\lambda_{min} (x) < {\lambda _{ij}}(x) <  \lambda_{max} (x)$ ,$x \in D$, e.g., we can let $\lambda_{min} (s) = \mathop {\inf }\limits_{s \le \left\| x \right\| \le r} ({\lambda _i}(x)),0 \le s \le r$, and $ \lambda_{max} (s) = \mathop {\sup }\limits_{\left\| x \right\| \le s} ({\lambda _i}(x)),0 \le s \le r,\forall i = 1,...,m$,$j=1,...,rank(J_i)$. It should be mentioned that if $D=R^n$, $r$ can be ignored.

Thus,  every diagonal element of $J(x) - \lambda_{min} (x)I$ can be written as
${\hat \lambda _{ij}}(x) = {\lambda _{ij}}(x) - \lambda_{min} (x) > 0$.
By the property of the positive definite matrix ,  $J(x) - \lambda_{min} (x)I$ is positive definite , therefore, we have $J(x) > \lambda_{min} (x)I$ .Similarly, we obtain $J(x) < \lambda_{max} (x)I$ . Combing two inequalities yields
\begin{equation}  \label{70}
\lambda_{min} (x)I < J(x) < \lambda_{max} (x)I
\end{equation}
Multiply  $P(x)$and $P^{-1}(x)$  on the both sides of  (\ref{70}),and define $\lambda_1 (x) = \lambda_{min}(x)$ and $\lambda_2 (x) = \lambda_{max}(x)$  yields
$ \lambda_1 (x)I < A(x) <  \lambda_2 (x)I $
Similarly, in the same way,  we can obtain above result when $A(x)$ is positive definite.

Based on above lemma, we have the following theorem.

\textbf{\emph{Theorem 2.}} Suppose system (\ref{27}) satisfies the assumption 2, if there exist  class $KL$ functions $\beta_{1i}< \beta_{2i},\forall i=1,...,n$  ,and a class $K$ differentiable function $\gamma$ such that for any initial state $x({t_0})$  and any bounded input $u(t)$ , the solution of $x(t)$ exists for all $t \ge {t_0}$  and satisfies
\begin{equation}\label{30}
\left\{ \begin{array}{l}
 {x_i}(t) \le {\beta _{2i}}({x_0},{u_0},t) + {\gamma _i}(u) - \int_0^t {{e^{ - \int_s^t {{\lambda _2}(\tau )d\tau } }}} \min (\gamma _i^{'}\dot u)ds \\
 {x_i}(t) \ge {\beta _{1i}}({x_0},{u_0},t) + {\gamma _i}(u) - \int_0^t {{e^{ - \int_s^t {{\lambda _1}(\tau )d\tau } }}} \max (\gamma _i^{'}\dot u)ds \\
 \end{array}\right.
\end{equation}
where $i = 1,...,n$,${\lambda_2}(u,e) \ge {\lambda_1}(u,e) > 0$.

\emph{Proof. } Define the error $e = x - \gamma (u)$ , then the error system of system (\ref{27}) can be written as
\begin{equation}  \label{31}
\dot e = f(e + \gamma (u),u) - {\gamma}^{'} {\dot u}
\end{equation}
where $\gamma {{'}} = \frac{{d\gamma (u)}}{{du}}$.

Expanding  $f(e + \gamma (u),u)$  yields
\begin{equation}\label{32}
{f(e + \gamma(u),u) = f(\gamma(u),u) + \nabla f^{'}(\xi  + \gamma (u),u)e}
\end{equation}
where $\xi  = \theta e,\theta  \in [0,1]$.
Due to $f(\gamma (u),u){\rm{ = 0}}$, then (\ref{31}) can be written as
\begin{equation}  \label{33}
\dot e = A(u,e)e - {\gamma^{'}}{\dot u}
\end{equation}
where $A(u,e) = \nabla {f^{'}}(\xi  + \gamma (u),u)$.

Because when $\dot u = 0$, system (\ref{27}) will converge to the constant $u$ ,  $\dot e = f(e + \gamma (u),u)$ is asympototiclly stable. By proposition 1, there exists Lyapunov function $V(e)$ such that
\begin{equation}  \label{34}
{\alpha _1}(\left\| e \right\|) \le V \le {\alpha _2}(\left\| e \right\|)
\end{equation}
\begin{equation}  \label{35}
\dot V = \frac{{\partial V(e)}}{{\partial e}}A(u,e)e \le  - W(e ),\forall \left\| e \right\| \ge \rho (\left\| \dot{u} \right\|)
\end{equation}
where  ${\alpha _1},{\alpha _2} \in {K_\infty }$ and $\rho  \in K$ ,  $W$ is a continuous positive definite function  on ${R^n}$ .

Calculating the derivative of (\ref{34}) yields
\begin{equation}  \label{36}
\frac{{\partial {\alpha _1}(\left\| e \right\|)}}{{\partial \left\| e \right\|}}\frac{{d\left\| e \right\|}}{{de}} \le \frac{{\partial V(e)}}{{\partial e}} \le \frac{{\partial {\alpha _2}(\left\| e \right\|)}}{{\partial \left\| e \right\|}}\frac{{d\left\| e \right\|}}{{de}}
\end{equation}
By the equivalence of the norms , there exists $k_1,k_2 > 0$  , such that for any norm ,the following equation is true
${k_2}{e^T}e \le \left\| e \right\| \le {k_1}{e^T}e$
Calculating the derivative of above equation  yields
\begin{equation}  \label{38}
{k_2}{e^T} \le \frac{{d\left\| e \right\|}}{{de}} \le {k_1}{e^T}
\end{equation}
In view of (\ref{38}), (\ref{36}) can be written as
\begin{equation}  \label{39}
\frac{{\partial {\alpha _1}(\left\| e \right\|)}}{{\partial \left\| e \right\|}}{k_2}{e^T} \le \frac{{\partial V(e)}}{{\partial e}} \le \frac{{\partial {\alpha _2}(\left\| e \right\|)}}{{\partial \left\| e \right\|}}{k_1}{e^T}
\end{equation}
Multiplying  $A(u,e)e$  to the both sides of (\ref{39}) yields
\begin{equation}  \label{40}
\frac{{\partial {\alpha _1}(\left\| e \right\|)}}{{\partial \left\| e \right\|}}{k_2}{e^T}A(u,e)e \le \frac{{\partial V(e)}}{{\partial e}}A(u,e)e \le \frac{{\partial {\alpha _2}(\left\| e \right\|)}}{{\partial \left\| e \right\|}}{k_1}{e^T}A(u,e)e,
\end{equation}
Consider (\ref{35}), we obtain
$\frac{{\partial {\alpha _1}(\left\| e \right\|)}}{{\partial \left\| e \right\|}}{k_2}{e^T}A(u,e)e \le  - W( e ).$
Due to ${\alpha _1} \in K$ ,$\frac{{\partial {\alpha _1}(\left\| e \right\|)}}{{\partial \left\| e \right\|}}{k_2} > 0$ , then for any $e \in {R^n}$,  ${e^T}A(u,e)e < 0$, that is  $A(u,e)$is negative definite .

By Lemma 1, there exist scalar functions ${\lambda _1}(u,e) > {\lambda _2}(u,e) > 0$ such that
$ - {\lambda _1}(u,e)I < A(u,e) <  - {\lambda _2}(u,e)I$.
Then there exists derivative inclusive
\begin{equation}  \label{41}
\dot e = {A_s}(u,e)e - {b_s}(u,\dot u)
\end{equation}
where ${A_s}(u,e) \in \{  - {\lambda _1}(u,e)I, - {\lambda _2}(u,e)I\} $,${b_s}(u,\dot u) \in \{ \min (\gamma _i^{'}{\dot u}),\max (\gamma _i^{'}{\dot u})\} $.

Solving (\ref{41}) yields
\[{e_i}(t) \le {e^{ - \int_0^t {{\lambda _2}(u,s)ds} }}{e_{i0}} - \int_0^t {{e^{ - \int_s^t {{\lambda _2}(u,e)d\tau } }}} \min (\gamma _i^{'}{\dot u})ds\]
\[{e_i}(t) \ge {e^{ - \int_0^t {{\lambda _1}(u,s)ds} }}{e_{i0}} - \int_0^t {{e^{ - \int_s^t {{\lambda _1}(u,e)d\tau } }}} \max (\gamma _i^{'}{\dot u})ds\]
In view of $e = x - \gamma (u)$   and define  ${\beta _{1i}}({x_0},{u_0},t) = {e^{ - \int_0^t {{\lambda _1}(u,e)ds} }}{e_{i0}}$ and ${\beta _{2i}}({x_0},{u_0},t) = {e^{ - \int_0^t {{\lambda _2}(u,e)ds} }}{e_{i0}}$,

we obtain
\[{x_i}(t) \le {\beta _{2i}}({x_0},{u_0},t) + {\gamma_i}(u) - \int_0^t {{e^{ - \int_s^t {{\lambda _2}(\tau )d\tau } }}\min (\gamma_i^{'}{\dot u})ds} \]
\[{x_i}(t) \ge {\beta _{1i}}({x_0},{u_0},t) + {\gamma _i}(u) - \int_0^t {{e^{ - \int_s^t {{\lambda _1}(\tau )d\tau } }}} \max (\gamma_i^{'}{\dot u})ds\]
\[\forall i = 1,...,n\]

\textbf{\emph{Remark 8}}. Since $\gamma (u)$ is the equilibrium point of (\ref{27}), $x(t)$ can converge towards $\gamma (u)$  at last when $\dot u \to 0$ , which is  different with theorem 1. The following two cases should be noted. If $x \in R$ , (\ref{30}) can be written as
\begin{equation}  \label{48}
x(t) = \beta ({x_0},{u_0},t) + \gamma (u) - \int_0^t {{e^{ - \int_s^t {{k}(\tau )d\tau }}}} {\gamma^{'}}\dot uds
\end{equation}
If  ${\gamma _i}(u) = {\gamma _j}(u),\forall i,j = 1,...,n$,(\ref{30}) can be written as
\[\left\{ \begin{array}{l}
 {x_i}(t) \le {\beta _{2i}}({x_0},{u_0},t) + {\gamma _i}(u) - \int_0^t {{e^{ - \int_s^t {{\lambda _2}(\tau )d\tau } }}} \gamma _i^{'}{\dot u}ds \\
 {x_i}(t) \ge {\beta _{1i}}({x_0},{u_0},t) + {\gamma _i}(u) - \int_0^t {{e^{ - \int_s^t {{\lambda _1}(\tau )d\tau } }}} \gamma _i^{'}{\dot u}ds \\
 \end{array} \right.\]

\textbf{\emph{Example 3.}}
Some simple examples are  ${{\dot x}_1} =  - \tan x_1^{} + u $ and
\begin{equation}
 {{\dot x}_1} =  - 3x_1^3 + 3x_2^3 + u \\
 {{\dot x}_2} =  - 2x_2^3 + 2x_1^3 \\
\end{equation}

If let system (\ref{27}) be  a scalar system, we have
\begin{equation}  \label{42}
\dot x = f(x,u)
\end{equation}
where $x \in R$ , $f:R \times R \to R$ is the continuous and local Lipschitz function w.r.t. $x$ and $u$ .

Then like the corollary 1,we have the similar result.

\textbf{\emph{Corollary 2}}. Suppose the condition of theorem 2 is satisfied, the solution of system (\ref{42}) can be written as
\begin{equation}  \label{ns}
x(t){\rm{ = }}\beta ({x_0},{u_0},t) + (1 - {e^{ - \int_\xi ^t {a(s)ds} }})\gamma (u(t))
\end{equation}
where $\xi  \in [0,t]$,$a(x) > 0$, and $\beta  \in KL$.

\emph{Proof. } Based on (\ref{48}) and following the way of Corollary 1£¬ we can prove the corollary 2.




\section{Stability Analysis of Interconnected Systems}

In this section, we analyze three typical kinds of interconnected systems via the approach proposed in the section III.

\subsection{ A New Proof of Small-gain Theorem}

Consider the following system

x-subsystem:
\begin{equation}  \label{45}
\dot x = {f_1}(x,z)
\end{equation}

z-subsystem:
\begin{equation}  \label{46}
\dot z = {f_2}(z,x)
\end{equation}
where $x \in {R^{{n_1}}}$,$z \in {R^{{n_2}}}$ , $f_1$ and $f_2$ are continuous functions similar with  (\ref{2}).

x-subsystem and z-subsystem satisfy the  proposition 1 and the following assumption.

\textbf{\emph{Assumption 3. }}The gain ${\gamma _x}$ of x-subsystem is defferentiable. ${\gamma _x}$ and the gain ${\gamma _z}$ of z-subsystem satisfies
${\gamma _z} \circ {\gamma _x}(s) < s,\forall s \in [0,\infty ].$

Then we have the following theorem.

\textbf{\emph{Theorem 3.}} If the nonlinear system (\ref{45})(\ref{46}) are ISS and satisfy the assumption 3. Then the interconnected system is asymptotically stable.

\emph{Proof. }\textbf{Step 1.} Transform the interconnected system into the cascade system.

According to proposition 1, for z-subsystem there exists the  Lyapunov function $V(z)$ such that
\begin{equation}  \label{47}
{\alpha _{1z}}(\left\| z \right\|) \le V \le {\alpha _{2z}}(\left\| z \right\|)
\end{equation}
\[ \dot V \le  - {W_3}(z),\forall \left\| z \right\| \ge \rho_z (\left\| x \right\|) \]
where ${\alpha _{1z},\alpha _{2z}} \in {K_\infty }$ ,  $\rho_z  \in K$, and ${W_3}$  is the continuous positive definite function.

Consider equation (\ref{47}), the condition $\left\| z \right\| \ge \rho_z (\left\| x \right\|)$ can be strengthened as $V \ge {\alpha _{2z}}(\rho_z (\left\| x \right\|))$ .

For the  x-subsystem, by corollary 1, we have
\begin{equation}  \label{49}
\left\| x \right\| \le \beta ({x_0},{z_0},t) + \alpha _{1x}^{ - 1}{\rm{((1}} - {e^{ - \int_\xi ^t {k(s)ds} }}){\alpha _{2x}}(\rho_x (\left\| z \right\|))).
\end{equation}
where $k(t)>0$.

Substituting above equation into $V \ge {\alpha _{2z}}(\rho (\left\| x \right\|))$ yields
\[V \ge {\alpha _{2z}}(\rho_z (\beta ({x_0},{z_0},t) + \alpha _{1x}^{ - 1}(({\rm{1}} - {e^{ - \int_\xi ^t {k(s)ds} }}){\alpha _{2x}}(\rho_x (\left\| z \right\|))))\]
Consider equation (\ref{47}), above equation is strengthened as
\begin{equation}  \label{50}
\left\| z \right\| \ge \alpha _{1z}^{{\rm{ - 1}}} \circ {\alpha _{2z}} \circ \rho_z (\beta ({x_0},{z_0},t) + \alpha _{1x}^{ - 1}(({\rm{1}} - {e^{ - \int_\xi ^t {k(s)ds} }}){\alpha _{2x}}(\rho_x (\left\| z \right\|))),
\end{equation}
Due to $\beta ({x_0},{z_0},t) \in KL$,(\ref{50}) is equivalent to
\begin{equation}  \label{51}
\left\| z \right\| > \alpha _{\rm{1}}^{{\rm{ - 1}}} \circ {\alpha _2} \circ \rho (\alpha _1^{ - 1}(({\rm{1}} - {e^{ - \int_\xi ^t {k(s)ds} }}){\alpha _2}(\rho (\left\| z \right\|)))
\end{equation}
Due to $({\rm{1}} - {e^{ - \int_\xi ^t {k(s)ds} }}) \in ({\rm{0}},{\rm{1)}}$ and by the monotonically increasing property of  class  $K$ function, we have
$\alpha _{1z}^{{\rm{ - 1}}} \circ {\alpha _{2z}} \circ \rho_z (\alpha _{1x}^{ - 1}({\alpha _{2x}}(\rho_x (\left\| z \right\|)))
  > \alpha _{1z}^{{\rm{ - 1}}} \circ {\alpha _{2z}} \circ \rho_z (\alpha _{1x}^{ - 1}(({\rm{1}} - {e^{ - \int_\xi ^t {k(s)ds} }}){\alpha _{2x}}(\rho_x (\left\| z \right\|)))$.
  That is
${\gamma_z}^\circ {\gamma_x}(\left\| z \right\|) > \alpha_{2z}^{ - 1} \circ {\alpha_{2z}}^\circ {\rho_z}(\alpha_{1x}^{ - 1}(({\rm{1}} - {e^{ - \int_\xi ^t {k(s)ds} }}){\alpha_{2x}}({\rho_x}(\left\| z \right\|))).$
So $\dot V \le  - {W_3}(z),\forall \left\| z \right\| > {\gamma _z} \circ {\gamma _x}(\left\| z \right\|)$.

Then the original interconnected system is transformed into the following cascade form
\begin{equation}\label{52}
\left\{ \begin{array}{l}
 \dot V \le  - {W_3}(z),\forall \left\| z \right\| > {\gamma _z} \circ {\gamma _x}(\left\| z \right\|) \\
 \dot x = {f_1}(x,z) \\
 \end{array} \right.
\end{equation}
\textbf{Step 2.}  Analyze the stability  of the feedback loop.

By the assumption 2, there exists $\left\| z \right\| > {\gamma _z} \circ {\gamma _x}(\left\| z \right\|)$ , such that z-subsystem is stable .

\textbf{Step 3}. Analyze the stability of cascade system.

By the lemma 4.7 in [10]  and the ISS property of x-subsystem , the cascade system (\ref{52}) is stable, so is the interconnected system (\ref{45})(\ref{46}).

\subsection{ A Special Interconnected System}

If at least one of subsystems in an interconnected system is not ISS, the small-gain theorem is not used directly, e.g., the following system.

x-subsystem:
\begin{equation}  \label{53}
\dot x = {f_1}(x, - z)
\end{equation}

z-subsystem:
\begin{equation}  \label{54}
\dot z = {f_2}(t,x)
\end{equation}
where $x \in R$, $z \in R$, $f_1$and $f_2$ are continuous functions, x-subsystem  is ISS and z-subsystem satisfies the following  assumption£º

\textbf{\emph{Assumption 4.}}~~$\dot z = {f_2}(t, - k(t)\rho (z))$
is stable , where $z \in D$,$1 > k(t) > 0$ and $\gamma  \in K$ is the differentiable gain function of x-subsystem .

There exists the following theorem.

\textbf{\emph{Theorem 4. }}If the nonlinear system (\ref{53})is ISS and (\ref{54}) satisfies the assumption 4. Then the interconnected system is asymptotically stable in $D$.

\emph{Proof.} \textbf{Step 1.} Transform the interconnected system into the cascade system.

By corollary 2,  equation  (\ref{53}) can be written as
\begin{equation}  \label{55}
x(t){\rm{ = }}\beta ({x_0},{u_0},t) - (1 - {e^{ - \int_\xi ^t {a(s)ds} }})\gamma (z(t))
\end{equation}
where  $\gamma  \in K$ is the gain of (\ref{53})and $a(t)>0$. Taking the Lyapunov function $V(z)$ for (\ref{54}) yields
\begin{equation}  \label{56}
\dot V = \frac{{\partial V}}{{\partial z}}{f_2}(t,x)
\end{equation}
Substituting (\ref{55}) into (\ref{56}) yields the cascade system as follows
\begin{equation}  \label{66}
\left\{ \begin{array}{l}
 \dot V = \frac{{\partial V}}{{\partial z}}{f_2}(t,\beta ({x_0},{u_0},t) - (1 - {e^{ - \int_\xi ^t {a(s)ds} }})\gamma (z(t))) \\
 \dot x = {f_1}(x, - z)
 \end{array} \right.
\end{equation}

\textbf{Step 2.}  Analyze the stability  of the feedback loop.

By the theorem 3.4 in [10] , ignore $\beta ({x_0},{u_0},t)$, we have
$\dot V = \frac{{\partial V}}{{\partial z}}{f_2}(t, - (1 - {e^{ - \int_\xi ^t {a(s)ds} }})\gamma (z(t)))$.
By the assumption 3 and $1 > 1 - {e^{ - \int_\xi ^t {a(s)ds} }} > 0$ , z-subsystem   is stable.

\textbf{Step 3.} Analyze the stability of cascade system.

By the lemma 4.7 in [10]  and the ISS property of x-subsystem , the cascade system (\ref{66}) is stable, so is the interconnected system (\ref{53})(\ref{54}).


\textbf{\emph{Example 4.}}~~Other simple examples are in the following.

$\left\{ \begin{array}{l}
 {{\dot x}_1} = x_2^3 \\
 {{\dot x}_2} =  - (1 + x_1^2){x_1} - {x_2} \\
 \end{array} \right.$
and
$\left\{ \begin{array}{l}
 {{\dot x}_1} =  - {x_1} + x_2^3 \\
 {{\dot x}_2} =  - {x_1} \\
 \end{array} \right.$

\subsection{The Consensus Analysis of Multi-agent Systems-A Simple Case}

Traditionally,  Lyapunov function based method is the popular tool to analyze the consensus. Here, we will present another way from the structural perspective.In order to illustrate how to  use the method proposed in section III to analyze the consensus problem, we consider the following nonlinear multi-agent system .

x-subsystem:
\begin{equation}  \label{57}
\dot x = {(z - x)^3}
\end{equation}

z-subsystem:
\begin{equation}  \label{58}
\dot z = {(x - z)^3}
\end{equation}
where  $z,x \in R$,suppose $\gamma_x$ and  $\gamma_z$  are the gains of the x-subsystem and z-subsystem respectively.

Even though  every system is ISS, the composition of the gain $\gamma_x(s) \circ \gamma_z(s) = s$
  causes the small-gain theorem is not applied directly in the situation. Following the method proposed in this paper, we have the following steps.

\textbf{Step 1.} Transform the interconnected system into the cascade system.

By the theorem 2,  the solution of x-subsystem is
\[x(t) = \beta ({x_0},{u_0},t) + z - \int_0^t {{e^{ - \int_s^t {a(\tau )d\tau } }}} \dot zds\]
where $a(t)>0$.
Substituting above equation into  z-subsystem yields
\begin{align}
 \dot z &= L(z,t)(\beta ({x_0},{u_0},t) - \int_0^t {{e^{-\int_s^t {a(\tau )d\tau } }}} \dot zds) \nonumber\\
 \end{align}
where
$L(z,t) = {(\beta ({x_0},{u_0},t) - \int_0^t {{e^{-\int_s^t {a(\tau )d\tau } }}} \dot zds)^2}.$
Then the original system is transformed into the following cascade system
\begin{equation}\label{63}
\dot z = L(z,t)\beta ({x_0},{u_0},t) - L(z,t)\int_0^t {{e^{-\int_s^t {a(\tau )d\tau } }}} \dot zds
\end{equation}
\begin{equation}\label{64}
 \dot x = {(z - x)^3}
\end{equation}

\textbf{Step 2.}  Analyze the stability  of the feedback loop.

For z-subsystem,  define $q(t) = {e^{\int_0^t {a(s)ds} }}{e^{\int_0^t {L(z,s)} ds}}$
Multiplying both sides of (\ref{63}) by  $q(t)$ and rearrange the equation we obtain
\begin{align}  \label{59}
&q(t)\dot z + q(t)L(z,t)\int_0^t {{e^{-\int_s^t {a(\tau )d\tau } }}} \dot zds \nonumber\\
&= q(t)L(z,t)\beta ({x_0},{u_0},t),
\end{align}

From the expression of  $q(t)$, one can verify
\begin{align} \label{62}
 & q(t)\dot z + q(t)L(z,t)\int_0^t {{e^{-\int_s^t {a(\tau )d\tau } }}} \dot zds \nonumber\\
  &= \frac{d}{{dt}}({e^{\int_0^t {L(z,s)} ds}}\int_0^t {{e^{\int_0^s {a(\tau )d\tau } }}} \dot zds) 
 \end{align}
Using (\ref{62}) in (\ref{59}) and integrating both sides of (\ref{59}), we obtain
${e^{\int_0^t {L(z,s)} ds}}\int_0^t {{e^{\int_0^s {a(\tau )d\tau } }}} \dot zds = \int_0^t {q(s)L(z,s)\beta ({x_0},{u_0},s)ds} $.
Therefore,
\[\int_0^t {{e^{\int_0^s {a(\tau )d\tau } }}} \dot zds = \int_0^t {{e^{\int_0^s {a(\tau )d\tau } }}{e^{\int_s^t { - L(z,\tau )} d\tau }}L(z,s)\beta ({x_0},{u_0},s)} ds\]
By theorem 2,due to $\beta ({x_0},{u_0},t) = {e^{ - \int_0^t {a(s)ds} }}$ ,then
$\int_0^t {{e^{\int_0^s {a(\tau )d\tau } }}} \dot zds = \int_0^t {{e^{\int_s^t { - L(z,\tau )} d\tau }}L(z,s)ds} .$
We obtain
\[\dot z = L(z,t)\beta ({x_0},{u_0},t) - L(z,t){e^{ - \int_0^t {a(s)ds} }}\int_0^t {{e^{\int_s^t { - L(z,\tau )} d\tau }}L(z,s)} ds.\]
Since  $L(z,s) = \frac{d}{{ds}}(\int_s^t { - L(z,\tau )} d\tau )$ , we have
\[\int_0^t {{e^{\int_s^t { - L(z,\tau )} d\tau }}L(z,s)ds}  = {e^{\int_s^t { - L(z,\tau )} d\tau }}\left| {_0^t} \right. = 1 - {e^{\int_0^t { - L(z,\tau )} d\tau }}.\]
Due to  ${e^{\int_s^t { - L(z,\tau )} d\tau }} \in KL$, we have
$\int_0^t {{e^{\int_s^t { - L(z,\tau )} d\tau }}L(z,s)} ds \to 1 $ ,
and

${e^{ - \int_0^t {a(s)ds} }}\int_0^t {{e^{\int_s^t { - L(z,\tau )} d\tau }}L(z,s)}  \to 0$,$\forall t \to \infty $.
Namely, $\int_0^t {{e^{ - \int_s^t {a(\tau )d\tau } }}} \dot zds \to 0$, and $ L \to 0, \forall t \to \infty $.

Thus $\dot z \to 0,\forall t \to \infty $ .
Because $L(z,t)\beta ({x_0},{u_0},t) > 0$,
$\int_0^t {L(z,t)\beta ({x_0},{u_0},t)ds} $
is integrable, so is
$\int_0^t {L(z,t){e^{ - \int_0^t {a(s)ds} }}\int_0^t {{e^{\int_s^t { - L(z,\tau )} d\tau }}L(z,s)} ds} $
, therefore
\begin{align}
z(t) &= z(0) + \int\limits_0^t {L(z,t)\beta ({x_0},{u_0},t)} ds \nonumber\\
&- L(z,t){e^{ - \int_0^t {a(s)ds} }}\int_0^t {{e^{\int_s^t { - L(z,\tau )} d\tau }}L(z,s)} ds \nonumber\\
\end{align}
exists.

\textbf{Step 3.} Analyze the stability of cascade system.

By the theorem 2, for x-subsystem , when $z(t) \to c$ where $c$ is a constant, $x(t) \to z(t)$, $\forall t \to \infty $. Therefore, the system is consensus.


\section {CONCLUSION}
In this paper, based on the structural perspective, we propose a new framework for the stability analysis of interconnected systems. This method bases on a new formulation of ISS concept which  weakens the conservative of the original form  and can analyze some problems which  can not be done by the small-gain theorem. As its applications, we also investigates the three  kinds of typical interconnected systems respectively. It should be mentioned that we  use a unifying approach to treat three different problems and provide a deep insight for  the common essence of them.
Further more this method can be used to design the lower-order controller for the high-order minimum-phase systems e.g.\cite{Y. Wang2}\cite{Y. Wang3}\cite{Y. Wang4}. In the future, we will combine this method and the idea in \cite{Y. Wang1} to analyze the consensus of MAS with nonlinear protocol on any topology and with the time-delay in the communication.

\end{document}